\documentclass[twocolumn, prd]{revtex4}
\usepackage[latin1]{inputenc}
\usepackage{amsthm,amsfonts}
\usepackage{amsmath}
\usepackage{amsthm}
\usepackage{amssymb}
\usepackage{appendix}
\usepackage{enumerate}
\usepackage{graphicx}
\usepackage[all]{xy}
\usepackage{comment}
\usepackage{graphicx,color}

\begin{document}

\newtheorem{theorem}{Theorem}[section]
\newtheorem{lemma}[theorem]{Lemma}
\newtheorem{proposition}[theorem]{Proposition}
\newtheorem{corollary}[theorem]{Corollary}

\newenvironment{Proof}[1][Proof]{\begin{trivlist}
\item[\hskip \labelsep {\bfseries #1}]}{\end{trivlist}}
\newenvironment{definition}[1][Definition]{\begin{trivlist}
\item[\hskip \labelsep {\bfseries #1}]}{\end{trivlist}}
\newenvironment{example}[1][Example]{\begin{trivlist}
\item[\hskip \labelsep {\bfseries #1}]}{\end{trivlist}}
\newenvironment{remark}[1][Remark]{\begin{trivlist}
\item[\hskip \labelsep {\bfseries #1}]}{\end{trivlist}}

\newcommand{\Qed}{\nobreak \ifvmode \relax \else
      \ifdim\lastskip<1.5em \hskip-\lastskip
      \hskip1.5em plus0em minus0.5em \fi \nobreak
      \vrule height0.75em width0.5em depth0.25em\fi}

\newcommand{\BE}{\begin{equation}}
\newcommand{\EE}{\end{equation}}
\newcommand{\BEN}{\begin{equation*}}
\newcommand{\EEN}{\end{equation*}}
\newcommand{\Ga}{\alpha}
\newcommand{\Gb}{\beta}
\newcommand{\Gc}{\chi}
\newcommand{\Gd}{\delta}
\newcommand{\Ge}{\epsilon}
\newcommand{\Gf}{\phi}
\newcommand{\Gg}{\gamma}										
\newcommand{\f}[2]{\frac{#1}{#2}}
\newcommand{\p}{\partial}
\newcommand{\m}[1]{\mathcal{#1}}
\newcommand{\IBF}[1]{\textit{\textbf{#1}}}
\newcommand{\un}[2]{\underset{#1}{#2}}
\newcommand{\Ov}[1]{\overline{#1}}
\newcommand{\pU}[2]{\f{\p\Ov{x}^{#2}}{\p x^{#1}}}
\newcommand{\pD}[2]{\f{\p{x}^{#1}}{\p\Ov{x}^{#2}}}
\newcommand{\Le}{\left}
\newcommand{\Ri}{\right}
\newcommand{\Bra}[1]{\Le<{#1}\Ri|}
\newcommand{\Ket}[1]{\Le|{#1}\Ri>}
\newtheorem{Teo}{Teorema}	
\newtheorem{Lema}{Lema}	
\newtheorem{Def}{Definição}
\newtheorem{Postulado}{Postulado}
\newtheorem{Pres}{Prescription}
\newcommand{\mf}[1]{\mathfrak{#1}}
\newcommand{\Pg}{\m{P}_{\m{N}}}
\newcommand{\PgA}{\m{G}_P}
\newcommand{\Pl}{\mf{p}_{\m{N}}}
\newcommand{\act}{\rhd}
\newcommand{\C}[1]{#1}

\title{A conceptual problem for non-commutative inflation and the new approach for non-relativistic inflationary equation of state}
\author{U. D. Machado}
\email[]{uda2@cornell.edu}
\altaffiliation{Center for Radiophysics and
Space Research.
Cornell University,
613 Space Science Building
Ithaca, NY, 14853
 USA} 
\altaffiliation{Departamento de Astronomia, Geofísica e Ciências Atmosféricas, Universidade de São Paulo.  Rua do Matão, 1226-Cidade Universitária-São Paulo/SP- 05508-090}

\author{R. Opher}
\email[]{opher@astro.iag.usp.br}
\altaffiliation{School of Electrical and Computer Engineering-Cornell University, 
428 Phillips Hall
Ithaca, NY, 14853
 USA}
\altaffiliation{Departamento de Astronomia, Geofísica e Ciências Atmosféricas, Universidade de São Paulo.  Rua do Matão, 1226-Cidade Universitária-São Paulo/SP- 05508-090}

\begin{abstract}
In a previous paper, we connected the phenomenological non-commutative inflation of Alexander, Brandenberger and Magueijo (2003) and Koh S and Brandenberger (2007) with the formal representation theory of groups and algebras \C{and analyzed minimal conditions that the deformed dispersion relation should satisfy in order to lead to a successful inflation}. \C{In that paper, we showed that elementary tools of algebra allow a group like procedure in which even Hopf algebras (roughly the symmetries of non-commutative spaces) could lead to the equation of state of inflationary radiation}. In this paper, we show that there exists a conceptual problem with the kind of representation that leads to the fundamental equations of the model. \C{The problem comes from an incompatibility between one of the minimal conditions for successful inflation (the momentum of individual photons is bounded from above) and the group structure of the representation which leads to the fundamental inflationary equations of state. We show that such a group structure, although mathematically allowed, would lead to problems with the overall consistency of physics, like in scattering theory, for example.  Therefore, it follows that the procedure to obtain those equations should be modified according to one of two possible proposals that we consider here.} One of them relates to the general theory of Hopf algebras \C{while} the other is based on a representation theorem of Von Neumann algebras, a proposal already suggested by us to take into account interactions in the inflationary equation of state. This reopens the problem of finding inflationary deformed dispersion relations and all developments which followed the first paper of Non-commutative Inflation. \C{In particular, this analysis shows us that we really need to consider the Hopf algebra language (here explained at an elementary level) in order to adequately formulate non-commutative inflation, a point only heuristically suggested in the original papers.}   
\end{abstract}

\maketitle
\section{Introduction}

Non-commutative inflation, as originally conceived \cite{Noncommutative_Inflation}, is a phenomenological model of inflation inspired by a varying speed of light cosmology (\cite{VSLCP} and \cite{NVSLC}). It was proposed as a possible mechanism connecting non-commutative spaces with inflation. In this model, radiation, rather than a scalar field, drives inflation. Non-commutative inflation is an application of a conjecture that the non-relativistic character of non-commutative spaces \cite{LPN} and \cite{Carmona}, modeled by a non-relativistic dispersion relation, was enough to determine the thermodynamics of radiation.  

Despite the original constraint with non-commutative spaces, related to many interesting Cosmological applications like in \cite{Nicolini2} and \cite{Nicolini3}, non-commutative inflation is actually a bridge connecting general violations of Lorentz symmetry with inflation, an idea that can be related with a wider literature in high energy physics \cite{Doubly}, \cite{AmelinoGamma}, \cite{ViableLV}, \cite{LimVSL},  \cite{Remo} and \cite{Nicolini}.  Instead of the usual Lagrangian approach of physics, in which non-relativistic effects are modeled by non-relativistic terms that can affect the cosmological perturbations,  such as in the so called trans-Planckian problem of inflation \cite{Martin} and \cite{matrinDsf}, the connection between non-commutative spaces and the fundamental equations of non-commutative inflation can be more directly related with the Wigner approach for QFT and the idea of Hopf algebra deformations of the Poincare Lie algebra due to non-commutative spaces \cite{Hopf}, rather than the star product approach \cite{Douglas}. Although a general procedure to deform a Lie algebra as a function of non-commutative space information is not available in the literature, postulating the existence of a generalization of the analysis \cite{Hopf}, it is possible to define the quantum theory of matter, not by a quantization of a deformed action assumed to be invariant by Hopf algebra, as previously done, or the star product approach to non-commutative field theories \cite{Douglas}, but as a deformation of the Wigner procedure to define quantum theory in terms of representations of the Poincaré group as quantum symmetries \cite{Machado}. This procedure is assured, by the general techniques of algebra, like the GNS construction (see \cite{Neumann} and \cite{Strocchi}), to have much in common with a Lie group representation, in particular, the fundamental role of the dispersion relation in the scheme.  

The Hopf algebra concept, more than an effort to bridge non-commutative spaces with non-commutative inflation, is actually a key concept in a more rigorous formulation of non-commutative inflation, being a useful concept in group theory in terms of which one can ask what is the group (algebraic) structure related to inflation. Something similar to ask what is the algebraic structure related to a fundamental invariant length scale and a fundamental speed, leading to the seminal paper of the doubly special relativity  \cite{Doubly} (see a generalization of this idea to include an invariant expansion rate \cite{DSRCS}). More than that, the algebra is the key connecting a phenomenological deformed dispersion relation with the effective energy-momentum tensor of inflationary matter.
  
\C{
Our main conclusion in this paper is that the Fock space structure, the basis underlying non-commutative inflation, is incompatible with minimum duration of inflation. Also, we propose alternatives to this structure that should be applied in order to obtain new fundamental equations.}
  
\C{
The paper is organized as follows:}

\C{
In section (II) we sketch a Wigner like argument which could be (at first sight)  considered as a more formal proof that the fundamental equations of non-commutative inflation are actually a consistent way to connect deformed Poincaré symmetries with a particular kind of equation of state for radiation in the early universe.}
 
\C{  
In section (III) we introduce a useful concept for our analysis, the Hopf Algebra concept. This is a generalization of the concept of Lie Algebra which is the algebraic structure underlying the Poincaré group.  Previous works in the literature suggest that Hopf algebras play an important role in defining the meaning of symmetry in non-commutative spaces.
}

\C{ 
In section (IV), we introduce elementary tools of algebra which allow a connection between Hopf algebras and the familiar structure of physical quantum theories in Hilbert space. This, at first sight, suggests that there should not be any modification in the argument sketched in section (II). We then state an algebraic prescription to define Non-commutative inflation that is in harmony with the original equations. That is essentially a Fock space like structure written in group theory terms. 
}

\C{
In section (V), we show that if we want to consider a more general symmetry than Poincaré, minimal physical requirements will constrain the way in which we are allowed to do that. These are related to things like (among others):
\begin{itemize}
\item	 A system which is in a stable state according to one inertial frame should be in a stable state in all other inertial frames.\\
\item	If an observer observes a scattering experiment preserving energy and momentum in one reference frame, all other observers will agree that energy and momentum are conserved by scattering. 
\end{itemize}
}

\C{     
  In section (VI), we recall a minimal condition for successful inflation which is a result of our previous paper. That is that the condition that zero mass one-particle states should have an upper bound to momentum. We then show that if we construct the more general representation of a symmetry group satisfying the conditions of the previous section, we find that the momentum of the theory itself must have an upper bound, including the classical macroscopic world. However, this excludes the prescription outlined in section (IV). This also implies that we are no longer dealing with Lie Algebras, but general Hopf algebras, and therefore strengthens the bond between non-commutative spaces and non-commutative inflation.  We argue that what happens here is similar to what happens in relativistic quantum theory in which representations with no lower bound to energy are mathematically possible but not physically allowed. It follows that the Fock space representation is mathematically allowed but not physically allowed.  
}

\C{  
In section (VII), we use concepts and methods of previous sections to propose two alternative ways  (which are consistent with the requirements of section (V)) to use the one-particle information to construct the multi-particle theory. Neither of them is a Fock space like structure. One of those can in principle be applied to an interacting case and will assure a well defined finite and unitary quantum theory with a low energy relativistic limit if we already have the relativistic counterpart. The equivalence of the two prescriptions or not is still not established.   These prescriptions still require closed form for the equation of state of the Non-commutative radiation. This is postponed for a future work.
We then present our conclusions in the last section.  
}

\section{A group theory argument for non-commutative inflation}

The basic idea explored in \cite{NVSL} and \cite{Noncommutative_Inflation} is that the effect of generic models of non-commutative space-time can be codified in the  modification of the energy-momentum relation. This is indicated in most of the implementations of the non-commutative principle to field theory \cite{LPN}, \cite{Carmona}. This would affect the calculation of the canonical partition function for radiation that in turn affects the early phases of the universe in thermal equilibrium.

This idea is formalized in the Wigner approach to relativistic quantum theory, as already discussed in \cite{Machado}, in which the basic problem is to construct representations of the Poincaré group as quantum symmetries, not the quantization of a particular classical field. That is, finding the correspondence $\pi$  among Lorentz transformations $x\to\Lambda x+a^\mu$, denoted by $(\Lambda,a^\mu)$, and the transformation $\pi[(\Lambda,a^\mu)]$: $\m{H}_{P}\to\m{H}_{P}$, that satisfies:
\begin{gather}
\pi[(\Lambda_1,a_1^\mu)]\cdot\pi[(\Lambda_2,a_2^\mu)]=\pi[(\Lambda_1,a_1^\mu)\cdot(\Lambda_2,a_2^\mu)],\label{R1}\\
\pi[(\Lambda_1,a_1^\mu)]^{-1}=\pi[(\Lambda_1,a_1^\mu)^{-1}],\label{R2}\\
\pi[(\Lambda,a^\mu)]\mbox{ is a quantum symmetry}\label{R3},
\end{gather}
where  $\m{H}_{P}$ is the projective Hilbert space, that is, the space of rays, $\cdot$ denotes the composition and $()^{-1}$ the inverse.  The symmetry is characterized by the invariance of transition amplitudes: 
\BE\label{Symmetry}
 |(T\Phi,T\Psi)|=|(\Phi,\Psi)|.
\EE

 According to the Wigner theorem, a quantum symmetry can be extended from rays in $\m{H}_P$ describing quantum states to the entire Hilbert space $\m{H}$ as a linear unitary or a antilinear antiunitary transformation. After such an extension, we might be dealing with projective representations. i.e. eq.(\ref{R1}) might be valid except for a phase $e^{i\theta}$;  
 
\C{It is implicit in the Wigner prescription to quantum symmetries that all transformations act in the same physical space, which is left invariant by them, all transformations have an inverse and the composition is also an allowed transformation, they therefore form a group. That is, despite all of the possible complex algebraic structures that one may consider, we are always dealing with groups.}

 The Poincaré group $P$ is a very particular group, a Lie group, that is, a group with topological structure. In addition, a Lie group which posesses a connected subgroup, the proper and orthochronous part $P^{\uparrow}_{+}$ (for which the Lorentz subgroup satisfies ${\Lambda^0}_0>0$ and $\det\Lambda=1$). The algebraic structure of the group, codified in the multiplication rule, does not constrain its topology and we could consider a particularly useful representative of this group, the so called universal enveloping group, in which the topology is simply connected, that is, every path between two points in the group space can be deformed continuously to any other. \C{For the Poincaré group, this group is $SL(2,\m{C})$}. For this group, one can always adjust phases in every $\pi[(\Lambda_1,a_1^\mu)]$ in such a way that it will be a non-projective representation.  
 
For connected Lie groups, the symmetry representation problem reduces to construct unitary representations, since, at least in the neighborhood of the identity described by canonical coordinates, every group element is part of a one-parameter subgroup and thus is the square of some other element. In addition, every element of a connected Lie group can be written as a product of elements \C{in} an arbitrarilly small neighborhood of the identity.    The square of an antilinear antiunitary operator is linear and unitary. The continuous$^{\underline{1}}$ \footnotetext[1]{i.e. $\Bra{\Psi}U(\lambda)\Ket{\Psi}$ continuous for every $\Phi$ in the Hilbert space $\m{H}$} one parameter family of unitary (non-projective) transformations affords a further simplification \C{due to the Stone theorem}, that can be put in the form $e^{-iH \lambda}$, $H$ \C{being a} selfadjoint operator. Therefore, the information about the group, in the connected Lie group case, can be cast in terms of information about a set of generators $H_i$ of one parameter subgroups of the canonical coordinates in the neighborhood of identity.

The set of generators \C{forms} a linear Lie algebra:

\begin{gather}
[P^\mu,P^\nu]=0\label{HP}\\
[M^{\mu\nu},P^\lambda]=i\Le(g^{\mu\nu}P^\lambda-g^{\lambda\nu}P^\mu\Ri)\\
[M^{\mu\nu},M^{\rho\sigma}]=-i\Le(g^{\mu\rho}M^{\nu\rho}-g^{\nu\rho}M^{\mu\sigma}+g^{\nu\sigma}M^{\mu\rho}-g^{\mu\sigma}M^{\nu\rho}\Ri)\label{LieM}
\end{gather}

 A unitary representation of a group can be further simplified by a change of basis, in such a way that all matrices $\pi[(\Lambda_1,a_1^\mu)]$ can be put in the block diagonal form. If this process cannot be continued, every block is an example of the simplest kind of representation, the so called irreducible representation.  The unitary representation of a group summarizes  into constructing the irreducible pieces by which every other representation can be constructed by. These basic parts are identified with the Hilbert space of one particle states. The dispersion relation $\mathcal{C}(p)$ is the fundamental information in this process because it is a (self-adjoint) function of space and time translation generators which commutes with all other generators of the symmetry group (Casimir of the Lie algebra) and defines a bounded operator $^{\underline{2}}$\footnotetext[2]{An operator $A$ is bounded if $||A\Psi||\leq C ||\Psi|| $, with $C$ independent of $\Psi$} ($e^{iC(p)}$) which commutes with every element of the group.    
By an infinite dimensional version of Schur`s lemma, a unitary representation of a group is irreducible if, and only if, every bounded operator which commutes with every element of the group is a multiple of the identity.
Indeed, for the Poincaré group, the algebraically independent set of Casimir operators is: 
\begin{gather}
P^2=H^2-\vec{P}^2,\nonumber\\
W=-w^2,\mbox{ com } w_\rho=\f{1}{2}\epsilon_{\Ga\mu\nu\rho}P^\Ga M^{\mu\nu},\label{W}\\
sign(P^0)=\theta(P^2)\epsilon(P^0),\nonumber
\end{gather}

Given two irreducible representations of $SL(2,\m{C})$, $U_1$ and $U_2$ in the Hilbert spaces $\m{H}_1$ and $\m{H}_2$ respectively (with possible additional symmetries such as parity in the electrodynamics case), we can construct another representation in the tensor product Hilbert space $\m{H}_1\otimes\m{H}_2$ which describes two non-interacting relativistic particles: 
\BE \label{TensorP}
U_1\otimes U_2(\sum_{ij} c_{ij} \Psi_i\otimes \Phi_j)= \sum_{ij} c_{ij} (U_1\Psi_i)\otimes(U_2\Phi_j).
\EE
The generators of the Lie algebra are given by:
\BE
X_i^{\otimes}=X_i^{(1)}\otimes I + I \otimes X_i^{(2)},
\EE 
$X_i^{(i)}$ generators of the Lie algebra of the representation $U_i$. 
 
In particular, we can make $U_1=U_2$ and take into account indistinguishable particles by introducing a Hilbert space basis in the symmetrized form: 
\BE
S_N(\psi_1\otimes \cdots \otimes \psi_N)=\f{1}{N!}\sum_{\sigma} \psi_{\sigma(1)}\otimes\cdots \otimes \psi_{\sigma(N)},\quad \psi_i\in\m{H} 
\EE
where $\sigma$ is an index permutation, or in the antisymmetrized for:
\BE
A_N(\psi_1\otimes \cdots \otimes \psi_N)=\f{1}{N!}\sum_{\sigma}{sign(\sigma)} \psi_{\sigma(1)}\otimes\cdots \otimes \psi_{\sigma(N)},
\EE
where ${sign(\sigma)}$ is $1$ for an even permutation and $-1$ for an odd permutation. This defines two kinds of Hilbert spaces:
\BE
\m{H}_S^N=\{\psi\in \m{H}^N;S_N\psi=\psi\}
\EE
and
\BE
\m{H}_A^N=\{\psi\in \m{H}^N;A_N\psi=\psi\}.
\EE
$\m{H}^N$ denoting the tensor product of $N$ Hilbert spaces.
  
We then define the Fock space of radiation:
\begin{equation}\label{Fock}
U_F=\sum_{N=0}^{\infty} {}^{\oplus}\left(\m{U}^{\otimes N}_{\lambda}\right)_{S}(\m{P})
\end{equation}
where $\m{U}^{\otimes N}_{\lambda}$ is the tensor product of $N$ photon irreducible representations characterized by the set of eingenvalues $\lambda$, in particular  $H^2-P^2=0$. $U_F$ is defined in the Hilbert space:
 \begin{equation}
 \mathcal{H}=\sum_{N=0}^{\infty}{ }^{\oplus}\left({\m{H}_S^N}_{\lambda}\right)
 \end{equation}
${\m{H}_S^N}_{\lambda}$ being the symmetrized tensor product of the Hilbert spaces where the photon irreducible representations acts. $N=0$ corresponds to the trivial vacuum representation $\pi[(\Lambda_1,a_1^\mu)]=I$.

The associated Hamiltonian is:
\BE\label{FockHamiltonian}
H_F=\sum_{N=1}^{\infty} { }^{\oplus} (H\otimes I \cdots \otimes I+ I\otimes H\otimes \cdots \otimes I+\cdots+ I\otimes I\cdots \otimes H)_N,
\EE

In the irreducible representation, $H$ can be put a form in which the momentum $P^j$ is diagonal. Being an irreducible representation, all we need to specify is the dimension $d$ of the eigenspace associated with the eigenvalue $p^j$ that can be proved to be the same for all $p^j$ in irreducible representations: 
\BE\label{IrreducibleH}
 H=diag_d(E(p),E(p),\cdots E(p)),
\EE 
that denotes a $d\times d$ diagonal matrix, $E(p)$ being the photon dispersion relation. The dimension $d$ is determined by the dimension of the corresponding little group, that is, the group subspace that leaves the four-momentum of the irreducible representation invariant.  For $P^2=0$ and $P^\mu\neq 0$, we can choose the four-momentum $k=(1,0,0,1)$ and using the correspondence between the Lorentz group action and  $SL(2,\mathbb{C})$ action on $2\times 2$ hermitian matrix, in such a way that $k$ corresponds to the matrix $\bar{p}$ given by:
\BE\label{LGM}
\bar{p}=\left(\begin{array}{cc}
1&0\\
0&0
\end{array}\right),
\EE
that is leaved invariant by the action of the elements $A\in SL(2,\mathbb{C})$ given by ($\bar{p}'=A\bar{p} A^*$): 
\BE
\gamma_{\phi}=\left(\begin{array}{cc}
e^{i\phi}&0\\
0&e^{-i\phi}
\end{array}\right),\quad
\gamma_\eta=\left(\begin{array}{cc}
1&\eta\\
0&1
\end{array}\right),
\EE
$\gamma_{\phi}$, $\phi$ real (mod $2\pi$), is a representation of the rotation group in two dimensions $SO(2)$, while $\gamma_\eta$, $\eta$ is complex, the translation group. The associated little group is therefore $ISO(2)$ of the rotations and translations in $\mathbb{R}^2$;  This subgroup is associated with translations and only admits infinite dimensional representations (except the trivial one) associated with the continous spin representations. Since we do not have knowledge of infinite internal degrees of freedom for particles, we postulate that this subgroup is only mapped into the trivial representation (all elements go into the identity). The $SO(2)$ subgroup has one dimensional irreducible representations, but since the electrodynamics is invariant by parity, that connects opposite helicities,  the dimension of the little group is two.

The canonical partition function is then given by:
\BE\label{PartitionF}
Z(\beta,V)=Tr\left(e^{-\beta {H_F}}\right)
\EE    
In order to determine the dependence with volume, we impose periodic boundary conditions on the one parameter subgroups of space translations in the $x^j$ direction: 
\BE U_{x^j}(0)=U_{x^j}(L).\EE 
It follows that:
\BE
U(x_i)=\sum_n e^{-i\f{2\pi n x_i}{L}}\int_0^{L} \f{1}{L}e^{i\f{2\pi n y}{L}}U(y)dy=\sum_n e^{-i\f{2\pi  n x_i}{L}}E_n,
\EE   
where $E_n$ are mutually orthogonal projections and:
\BE
\lim_{n\to \infty}\sum_{n}E_n\Psi=\Psi.
\EE
This implies momentum quantization, since $U(t)=e^{-i\f{P^i x^i}{\hbar}}$, where 
\BE\label{Qmomento}
P=\sum_n n E_n
\EE 
The allowed values of momentum are:
\BE\label{AllowedP} 
p_j=\f{2\pi\hbar n_j}{L},
\EE
where $n_j$ an integer.

The idea of non-commutative inflation \cite{NCVSL} is to introduce the fenomenological modification 
\BE\label{FDR}
 E^2=P^2\to E^2=P^2f^2(E)
\EE
straight into (\ref{IrreducibleH}) and (\ref{FockHamiltonian}), modifying the calculation of (\ref{PartitionF}) 
and obtaining, from the usual thermodynamic relations: 

\begin{equation}\label{rho}
   \rho(E,T)=\frac{1}{\pi^2}\frac{E^3}{\exp{E/T}-1}\frac{1}{f^3}\left|1-\frac{Ef^{'}}{f}\right|
   \end{equation}
   \begin{equation}\label{p}
   p=\f{1}{\beta}\f{\p}{\p V}ln\Le(Z(\beta,V)\Ri)=\frac{1}{3}\int \frac{\rho(E,T)}{1-\frac{Ef^{'}}{f}}dE
   \end{equation}
   \begin{equation}\label{rhoT}
    \rho=-\f{1}{V}\f{\p}{\p\beta}ln\Le(Z(\beta,V)\Ri)=\int\rho(E,T)dE.
   \end{equation}
We consider here $c=k_B=\hbar=1$. To find an inflationary behavior, \cite{Noncommutative_Inflation} used the ansatz:
\BE\label{fE}
   f=1+(\lambda E)^{\alpha},
\EE

We extended this to a set of inequalities on $f(E)$ related to minimal requirements for inflation \cite{BrandenbergerRevInf}. We use this extension as a starting point of our argument.   

\section{The Hopf algebra concept}\label{Hopf}

A question addressed by us in \cite{Machado} is wheter or not the above analysis  follows unaffected under the non-commutative hypothesis, since this model is supposed to be a mechanism for inflation applicable to non-commutative spaces. Nonetheless, what is meant by a quantum theory in the non-commutative space is a matter of discussion in the literature \cite{CameliaCo}, \cite{Gamboa}, \cite{GamboaQM}, \cite{Hopf} and \cite{Spectral}, we therefore followed a suggestion, more adjustable to the Wigner approach, that the effect of non-commutative space-time  is changing the algebraic structure of the Poincaré Lie algebra \cite{Hopf}. This is something similar to quantizing a group, in which the algebraic structure of the Lie algebra is transformed into that of a Hopf algebra. 

The information which defines the non-commutative space is the $C^*$-algebra $^{\underline{3}}$ \footnotetext[3]{$C^{*}$-algebra is a linear vector space $\mathcal{A}$ with an associative product $\cdot:\mathcal{A}\times\mathcal{A}\to\mathcal{A}$ (i.e. $(a\cdot b)\cdot c = a\cdot(b\cdot c) $); an operation called involution $*:\mathcal{A}\to\mathcal{A}$ that is defined  with the properties: $(A+B)^*=A^*+B^*$, $(\lambda A)^*=\overline{\lambda}A^*$, with $\lambda$ a complex number, $(AB)^*=B^*A^*$ and $(A^*)^*=A$ ;  a norm $||\mbox{ }||:\mathcal{A}\to\Re$  with respect to which the algebra is a Banach space (i.e. given a sequence $a_n$ of elements, if $\lim_{n\to\infty}||a_{n+m}-a_{n}||=0 $ for each $m>0$, then there exists an $a$ such that $\lim_{n\to\infty}||a_{n}-a||=0$ . );  the product is continuous with respect to the norm, i.e. $||AB||\leq||A||\cdot||B||$ , and the norm additionally  satisfies $||a^{*}a||=||a||^2$}, that, in some sense, is the algebraic idealization of a set of complex continuous functions defined on a topological space $X$. An important result, called the Gelfand-Naimark theorem, states that, when the product of the $C^*$-algebra is commutative, it is possible to recover the space $X$ as a unique function of the algebraic information alone. If we allow the product of the algebra to be non-commutative, we will no longer have an space $X$, but we can establish a non-unique correspondence with a set of Hilbert space operators. $C^*$-algebra is what one can call the flat non-commutative space, since the metric is not recovered by the algebraic information in this formalism. A generalization, that is the fundamental concept in the non-commutative geometry, is the spectral triple \cite{Spectral}, \cite{ST1},\cite{ST2},\cite{ST3},\cite{ST4} and \cite{ST5}, in which additional structure is furnished to the $C^*$-algebra to codify a metric.

Although the procedure (described in \cite{Hopf}) to take into account the information that defines the non-commutative space, the $C^*$-algebra, and deform the Lie algebra structure is not defined for general non-commutative spaces, if we assume that such a procedure must exist, we are able to define a formal deformation of the relativistic quantum theory in terms of algebraic methods \cite{Machado}. That is very similar to what is done in quantum field theory in curved space-time.  By doing this, we are following a different approach of the very authors of the Hopf algebra deformation procedure, in which, once possessing the associated Hopf algebra, they define the physics (only for real free scalar field) by a Lagrangian action to be quantized. A Lagrangian that is required to be invariant under Hopf Algebra action:
\BE\label{ActionHopf}
S[\phi(x)]=\int\phi(x)(\Box_\lambda-M^2)\phi(x),
\EE
where $\Box_\lambda$ is a non-relativistic differential operator.

The Hopf algebra concept, instead of a formal need to a possible connection with non-commutative spaces, is actually a fundamental ingredient of non-commutative inflation.

There are many ways to define a Hopf algebra, a possible way  which suggests its connections with non-commutative spaces is saying that the Hopf algebra is a set of transformations which acts not only on a vector space (like the Hilbert space), but on a algebra, like the $C^*$-algebra, and acts in a way dependent on the algebra product, that codifies non-commutativity.

A Hopf algebra $H$ acts in a algebra $\m{A}$  as a set of transformations not necessarily  invertible, $H:\m{A}\to\m{A}$, this action being denoted by $h\act f$, $h\in H$ and $f\in \m{A}$.  $H$ is an algebra, that is, one can multiply elements, and construct linear combinations with complex coefficients: 
\begin{gather}
(h_1\cdot h_2)\act f= h_1\act(h_2\act g)\label{HopfP}\\
(\Ga h_1+\Gb h_2)\act f= \Ga (h_1\act f)+\Gb (h_2\act f).
\end{gather} 
In this way, $H$ depends on the product of the elements of $\m{A}$ by a generalization of the Leibniz rule: $h\act (f\cdot g)=\sum_{i} (h_{i(1)}\act f)\cdot(h_{i(2)}\act g)$, or, by using a short notation:
\BE\label{CoproductP} 
h\act (f\cdot g)= (h_{(1)}\act f)\cdot(h_{(2)}\act g),
\EE
where the rule 
\BE\label{Coproduct} 
\Delta: {H}\to{H}\otimes{H}
\EE 
given by $\Delta: h\to\sum_{i}h_{i(1)}\otimes h_{i(2)}$  is called a \IBF{coprodut}.

It leads us to an alternative way to define a Hopf algebra, a way which is particularly important for the purposes in this paper. A Hopf algebra $H$ is an algebra which has a rule to construct tensor products that furnish representations of the same algebra. 

Consider for example a Lie Algebra realized in the Hilbert space $\m{H}$. Its generators satisfy:
\BE\label{LieAlgebra}
[X_i,X_j]=C_{ij}^kX_k
\EE
But we can define new representations of the very same Lie algebra (\ref{LieAlgebra}) in the tensor product of Hilbert spaces $\m{H}_1\otimes \m{H}_1$ by: 
\BE
\Delta X_i= X_i\otimes I+I\otimes X_i 
\EE
we observe that this is exactly what is done in (\ref{FockHamiltonian}).     

We can actually construct arbitrary tensor products, because $\Delta$ has a property called \IBF{coassociativity}: 
\BE
(id\otimes \Delta)\Delta h=(\Delta\otimes id)\Delta h
\EE

\xymatrix{ & h_{(1)}\otimes h_{(2)}\ar[rd]^{id\otimes \Delta}&\\ 
 h\ar[ru]^{\Delta\quad\quad}\ar[rd]_{\Delta\quad\quad} & &h_{(1)}\otimes h_{(2)}\otimes h_{(3)}  \\
  & h_{(1)}\otimes h_{(2)} \ar[ru]_{\Delta \otimes id}& \\
  }
 
 Here $id$ is the indentity among the operators  $H\to H$.  This follows from $H\act (abc)$, where we apply the rule (\ref{CoproductP})  first on the pair $a\cdot(bc)$, but it must yield the same result as applied first on the pair $(ab)\cdot c$.

A generalization of rule (\ref{CoproductP}) for the product of $N$ elements of $\m{A}$  follows from coassociativity:
\BE\label{CoproductG}
H\act (\prod_i f_i)= \prod_i (h_{(i)}\act f_i),
\EE 

In addition, we can define a generalization of the rule (\ref{Coproduct}) $\Delta^N:H\to H^{\otimes N+1}$: 
\BE\label{DeltaN}
\Delta^N h=(\Delta\otimes id\cdots \otimes id)\cdots(\Delta\otimes id\otimes id)(\Delta\otimes id)\Delta h,
\EE
which would yield, by coassociativity, the same result, even if in each term $(\Delta\otimes id\cdots \otimes id)$  we changed the positions of the $\Delta$. In this sense, we can identify the $\Delta$ itself with $(\Delta\otimes id\cdots \otimes id)$ while acting in $H^{\otimes N}$.  

If we have a rule $\Delta: H^{\otimes N}\to H^{\otimes N+1}$, we must have a rule to do the inverse. This rule is obtained when one of the multiplying elements of $\m{A}$, in the product rule (\ref{CoproductP}), is the identity of $\m{A}$: 
\begin{align*}
h\act (1\cdot g)&=(h_{(1)}\act 1)\cdot(h_{(2)}\act g)\\
                &=\epsilon(h_{(1)})\cdot({h_{(2)}\act g})\\
                &=h\act g;\\
h\act (g\cdot 1)&=(h_{(1)}\act g)\cdot(h_{(2)}\act 1)\\
                &=({h_{(1)}\act g})\cdot\epsilon(h_{(2)})\\
                &=h\act g;
\end{align*}
where the rule $\epsilon: H\to\mathbb{C}$ is called the \IBF{counity}. The counity undoes the $\Delta$ action, in the following sense: 
\xymatrix{ & \epsilon(h_{(1)})\otimes h_{(2)}\ar[rd]&\\ 
 \Delta h=h_{(1)}\otimes h_{(2)}\ar[ru]^{ \epsilon\otimes id}\ar[rd]_{id\otimes \epsilon} & &h  \\
  & h_{(1)}\otimes \epsilon(h_{(2)}) \ar[ru]& \\
  }
  
We therefore set $\epsilon: H^{\otimes N+1}\to H^{\otimes N}$, where $\epsilon$ is identified with $(\epsilon\otimes id\cdots \otimes id)$, independent of the position of $\epsilon$. 

The Hopf algebra $H$ acts in $\m{A}$ as a set of transformations not necessarily invertible. In spite of that, there is a generalized notion of inverse called antipode, $S:H\to H$. The antipode is such that if $H$ has an inverse, it will satisfy $S(h)=h^{-1}$. The antipode is (uniquely) defined by the properties $S(h_{(1)})\cdot h_{(2)}=h_{(1)}\cdot S(h_{(2)})=\epsilon(h)1_H$, where $1_H$ is the identity of $H$. In a short notation: 
\BE\label{AntipodaP} 
\cdot (S\otimes id)\Delta h=\cdot (id\otimes S)\Delta h=\epsilon(h)1_H.
\EE

 Given $h_{(1)}\otimes\cdots \otimes h_{(N+1)}=\Delta^N h$, we apply  $\cdot (S\otimes id)\otimes id \cdots \otimes id$, possibly changing the position of the term $\cdot(S\otimes id)$, which yields $\Delta^{N-2} h$. 

\section{Connecting Hopf algebras with the Hilbert space}

As already observed, in \cite{Hopf}, the authors suggest that, once  possessing a Hopf algebra which replaces the Poincaré Lie algebra, they define the physics (only for a real free scalar field) by an action functional to be quantized. An action functional that is required to be invariant under Hopf Algebra action:
\BE\label{ActionHopf}
S[h\act \phi(x)]=S[\phi(x)],
\EE
where $\phi(x)$ is an element of the non-commutative $C^*$-algebra which defines the non-commutative space. In our work, \cite{Machado}, we suggest an alternative formulation. This formulation consists in applying the Wigner prescription for relativistic quantum theory by assigning irreducible representations of the Hopf algebra to Hilbert spaces of one particle systems. A useful concept here is the GNS construction of the mathematical $C^*$ algebra theory.  

The GNS construction is of fundamental importance in physics and mathematics. It is the basis of the proof of the Von Neumann theorem, which states that all irreducible representations of the Heisenberg algebra are unitarily equivalent. The GNS construction is what assures us that the representation problem of $C^*$ algebras always has a solution, provided that all algebraic conditions of its definition are satisfied. It creates a correspondence between the algebra and a set of Hilbert space operators. 

The fundamental idea is that for all state $\omega:\m{A}\to \mathbb{C}$, i.e.  positive definite linear functional, defined by $\omega(A^*A)\geq 0$, corresponds a representation $\pi_\omega: \m{A}\to \m{O}(\m{H_\omega})$, $\m{O}(\m{H}_\omega)$ being operators in the Hilbert space $\m{H}_\omega$. It is done in such a way that $\m{H}_\omega$ has a cyclic vector $\Psi_\omega$, that is, the application, in $\Psi_\omega$, of the operators of the representation, generates the entire Hilbert space (more exactly, a dense subspace), in such a way that $\omega(A)=(\Psi_\omega,\pi_\omega(A)\Psi_\omega)$. Furthermore, any other representation $\pi$ in Hilbert space $\m{H}_\pi$ with a cyclic vector $\Psi$ which satisfies $\omega(A)=\Le(\Psi,\pi(A)\Psi\Ri)$ is unitarily equivalent to $\pi_\omega$, that is, there exists an unitary transformation $U:\m{H}_\pi\to\m{H}_\omega$ such that:
\BE
U\pi(A)U^{-1}=\pi_\omega(A),\quad U\Psi=\Psi_\omega.
\EE

The GNS construction is based on the fact that the $C^*$  algebra is already an Hilbert space, except for a scalar product to be defined. This can be taken as $(A,B)=\omega(A^*B)$, where $\omega$ is a state. The inner product is positive definite by positivity of the state, but $(A,A)=0$  does not imply  in general $A=0$, an inner product property. We define then the set: 
\BE
J=\Le\{A\in\m{A}, \omega(B^*A)=0, \forall B\in \m{A}\Ri\}
\EE
that is such that $\m{A}\cdot J\subseteq J$, property which makes $J$ a \IBF{left ideal}. The states of the Hilbert space are then equivalence classes defined in terms of $J$, that is, any two elements that differ by a $J$ element are equivalent: $[A]={A+B, B\in J}$, which defines the so called quotient space $\m{A}/J$. The Hilbert space operators are then given by: 
\BE
\pi_\omega(A)[B]=[AB].
\EE
and such that:
\BE
\omega(A)=\Bra{\Psi_\omega}A\Ket{\Psi_\omega}
\EE

The existence of representations is therefore assured by the existence of the states, positive definite linear functionals, but it is part of the $C^*$ algebra theory to prove that such states do exist, provided that the algebraic properties of the definition are satisfied.

Furthermore,  different kinds of states are related to different kinds of representations. In particular, the irreducible representations are related to normalized states (i.e. $\omega(1)=1$) that cannot be decomposed as convex linear combinations of any other two normalized states, i.e., $\omega(A)\neq \lambda \omega_1(A)+(1-\lambda) \omega_2(A)$, $\lambda\in[0,1]$ and $\omega_i$ an state, $\omega_i\neq \omega$. In particular, Schur's lemma is valid, and a state related to irreducible representation satisfies $\omega(P(C))=P(\omega(C))$, where $C$ is a Casimir of the algebra and $P(C)$ in a polynomial in $C$.

 We can do the same thing with Hopf algebras. There exists many ways in which a Hopf algebra can act on other algebraic structures and many ways to realize its action. The particular case given by Eq. (\ref{CoproductP}) is called a left corregular action and it is the particular way explored in \cite{Hopf} to define the Poincaré Lie algebra deformation under non-commutative hypothesis. In group theory, for example, the left multiplication of an element $g$ by another element $g'$ furnishes a representation of the group: $\pi[g](g')=g\cdot g'$, but it is not the case in Hopf algebras. To do that we must consider the linear functionals on it, the so callled dual Hopf algebra, $H^*: H\to \mathbb{C}$, denoted by $\left<\phi,h\right> \to \mathbb{C}$, $\phi\in H^{*}$, $h\in H$. We can transform $H^*$ into a Hopf algebra through the structure of $H$:  $\left< \phi\psi,h\right>=\left< \phi\otimes\psi,\Delta h\right>$, $\left< \Delta\psi,g\otimes h\right>=\left< \psi,g\cdot h\right>$, $\left< 1,h\right>=\epsilon(h)$ and $\epsilon(\psi)=\left<\psi,1\right>$.

We then define the action of $H$ into $H^*$ according to (\ref{CoproductP}), which transforms $H^*$ into the so called $H$-module algebra, yielding the so called left corregular representation $R^*:H^*\to H^*$: \BE 
\left<R^*_{g}(\phi), h\right>=<\phi,hg>=<\phi_{(1)},h>\cdot<\phi_{(2)},g>.
\EE 

To proceed in the GNS construction, we need to define a involution $*: H\to H$, the adjoint operation of $C^*$ algebras, characterized by the properties  $(A+B)^*=A^*+B^*$, $(\lambda A)^*=\overline{\lambda}A^*$ such that $(AB)^*=B^*A^*$ and $(A^*)^*=A$. We need actually to assure that the self adjoint property $X=X^*$ is compatible with the algebra in such a way that we can transform it into generators of one parameter subgroups. Furthermore, we need compatibility with the Hopf algebra structure in the following sense: $\Delta h^{*}=(\Delta h)^{*\otimes*}$, $\epsilon(h^*)=\overline{\epsilon(h)}$ and $(S\circ*)^2=id$. It assures things like, if $H$ is a self-adjoint operator, $\Delta H$ will be a self-adjoint operator too.   

As in the $C^*$ algebra case, $H^*$ is a Hilbert space, except for a inner product to be defined. This product must be such that $(\phi,h\act\psi)=(h^*\act \phi,\psi)$ $\phi,\psi\in H^*$. The Hopf algebra theory yields a natural choice of inner product, called the right integral  $\int:H^*\to \mathbb{C}$ defined by the property: 
\BE\label{RightInt}
(\int \otimes id)\circ \Delta=  1_{H^*}\cdot \int.
\EE
Analogously we can define the left integral, both being unique, except by a multiplicative factor. 

The importance of the right integral is how it is related to the left corregular action:
\BE\label{RIntp}
\int (\phi\act g)^* f=\int g^*(\phi^*\act f) 
\EE
We can define the ideal $J_{H^*}$:
\BE
J_{H^*}=\{f\in H^*/ \int g^* f=0, \forall g\in H^* \}
\EE
that defines the Hilbert space  $H^*/J_{H^*}$. It follows that we transform the Hilbert space into a non-commutative algebra in which the Hopf algebra acts following the rule (\ref{CoproductP}), the same rule as in it acts on the non-commutative $C^*$-algebra. 

We could, however, interpret the Hopf algebra simply as an algebra which can be realized as a $C^*$-algebra, by functionals that act on $H$ itself, not in $H^*$, that is, ignoring the rule (\ref{CoproductP}) and worrying about the coproduct only if we have to deal with tensor products. This is going to be the point of view of the rest of this work.  

The prescription to define the physics of the radiation that would lead to the same equations of the non-commutative inflation, (\ref{rho}) and (\ref{p}), would be:

\begin{Pres}[$\pi_{\mf{g}}$]\label{Prescricao1}
Be $\Pl$ a Hopf algebra with generators $X_i$ which deforms Poincaré Lie algebra $^{\underline{4}}$ \footnotetext[4]{More exactly, its universal enveloping algebra, where multiplication between elements is defined} with a set of Casimir elements $C_0,C_1\cdots$ and be $\pi_{\lambda'}:\Pl\to O(\m{D})$ the irreducible representation $^{\underline{5}}$ \footnotetext[5]{Associated to the GNS construction given by states in $\Pl$ or $\Pl^*$. $\m{O}(\m{D})$ denotes operators in the domain $\m{D}$.} with Casimir eigenvalues $\lambda'=(\lambda_0,\lambda_i)$.  Be  $\Pg$ the group generated by elements of the form $e^{-iX_i t}$ and  $\mathcal{U}_{\lambda'}^{\mathcal{NP}}$ an irreducible representation of it given by $\pi_{\Pg}[e^{-iX_i t}]=e^{-i\pi_{\lambda'}(X_i) t}$, $t\in \mathbb{R}$. The physics of the radiation is defined by a representation of $\Pg$  obtained by (\ref{Fock}) by the replacement:
\BE\label{FreeHopfPres}
\m{U}_{\lambda}\to \mathcal{U}_{\lambda'}^{\mathcal{NP}}, 
\EE
for some set $\lambda'$. This prescription leads us to the following representation of $\Pg$:
\BE\label{FockDef}
U^{\mathcal{NP}}_F=\sum_{N=0}^{\infty} {}^{\oplus}\left(\m{U}^{\mathcal{NP}\otimes N}_{\lambda'}\right)_{S}
\EE
\end{Pres} 

Observe that it is a Fock space structure, the difference is the mathematical structure of one-particle states. Nonetheless, if there are some things in common between the new one-particle structure and the usual relativistic one, equations (\ref{rho}) and (\ref{p}) will be the same.  

Also, observe that the role of the GNS construction, if not directly applied to realize the algebra, is going to be telling us that, once we define a $C^*$ (or Hopf) algebra  with all algebraic requirements, it is assured that the representation problem has a solution and we are able to do definitions like the above one. This suggests that we should define the group by definig a $C^*$ algebra with appropriate requirements for its realization.  
\section{ Important physical requirements on the Poincaré Lie algebra deformation $\Pl$ and on the representation of the associated group $\Pg$ that drives inflation }\label{CAlS} 

\subsection{Abstract algebraic conditions}
We thus define a deformation of the Poincaré Lie group $\m{P}$ through a deformation of its universal enveloping algebra $^{\underline{6}}$ \footnotetext[6]{The associative algebra of generators, that is, there exists an associative product such that the commutator is the Lie product} $\mf{p}$ to select an representation of it to describe the radiation and drive the inflation. This deformation, denoted by $\Pl$, besides the conditions to assure the existence of its realizations, must satisfy some physical requirements that we now describe. 

The first one is that $\Pl$ must form a nontrivial algebra, that is, its generators must satisfy a set of algebraic equations not trivially satisfied by every set of Hilbert space operators. It may appear a trivial condition, \C{since a thing that we usually take for granted is that, no matter what is the group we are talking about, the generators will always for an algebra. Nevertheless, that is not mathematically true. Physics cannot be formulated without an underlying algebraic structure for symmetry generators though. It is in terms of those nontrivial algebraic equations that the physical requirements are expressed. On the other hand, such a specification tell us what is the particular $\Pl$ that we are considering.} 

Furthermore, the generators of this algebra and the algebraic constraints must be compatible with the self adjoint condition $X=X^*$, for an involution satisfying  $(A+B)^*=A^*+B^*$, $(\lambda A)^*=\overline{\lambda}A^*$ and such that $(AB)^*=B^*A^*$. 

\C{A physical requirement related to the possibility of measure Energy and Momentum at same time, or the very concept of dispersion relation will not make sense, is that} $\Pl$ must have a commutative subalgebra associated to space-time translation generators. We will associate this algebra with the energy $E$ and momentum $P$. Besides, the algebra must have an Casimir $C(E,P)$ obtained as an algebraic function of $E,P$ that is compatible with the positive energy condition. 

Such compatibility is in the following sense: the solutions of the equation $C(E,P)=\lambda$ define surfaces in the $(E,P)$ variables, that we call a \IBF{mass-shell deformation}, denoted by $\m{C}_\lambda$, such that for some subset $M_{\Pl}$ of $\lambda$, $\m{C}_\lambda$ has a connected component satisfying $E\geq 0$ and $E=0\Leftrightarrow P^\mu=0$. 

We are going to call the set of the $(E,P)$ points such that $C(E,P)\in M_{\Pl}$ and $E\geq0$ the \IBF{deformed forward light cone} $V^+_{\Pl}$.  This property is part of the conditions that will assure that there are irreducible representations of $\Pg$ with positive-definite energy.   

The algebra generated by energy and momentum, denoted by $\mf{t}$, must be an \IBF{ideal} in the Lie algebra. That is, for any generator $X_i$, $[P^\mu,X_j]=G^\mu(P^\nu)$, where $G^\mu(P^\nu)$ is an algebraic function of energy and momentum. This condition assures that energy eigenstates, related to stable states, are mapped into energy eingenstates by the action of $\Pg$. That is, if the system is observed to be in a stable state in one reference frame, it will be observed to be in a stable state after the application of $\Pg$ $^{\underline{7}}$\footnotetext[7]{\label{IdealFootnote} Suppose $\Psi$ is an energy eingenstate, since $[E,P]=0$, it is also a momentum eigenstate. We can then write $P^\mu \Psi=p^\mu \Psi$, or, equivalently, $e^{iP^\mu} \Psi=e^{ip^\mu} \Psi$. A referential change by $\Pg$ implies $e^{iP^\mu}\to Ue^{iP^\mu} U^{-1}=e^{iC} e^{iP^\mu}$, what implies that $e^{iC}= Ue^{iP^\mu}U^{-1}e^{-iP^\mu}$. We can write $U=e^{iX_i}$, and apply the integral form of the Baker-Hausdorff formula: 
\begin{gather*}
e^{C}=e^{A}e^{B}\\
C=B+\int_0^1 g(e^{tad A}e^{ad B})A dt;\quad g=ln(z)/(z-1),
\end{gather*} 
where $ad A$ is the adjoint action given by $ad A(B)=[A,B]$. We conclude that $C\in\mf{t}$ and, furthermore, $e^{iC}e^{iP^\mu}=e^{i \bar{P}^\mu}$, where $\bar{P}^\mu$ is an analytic function  $P^\mu$ that defines a representation.}. The fact that the subalgebra $\mf{t}$ is an ideal essentially assures that:
\BE\label{RPnP}
U(\mf{g}_1)P^\mu U^{-1}(\mf{g}_1)= G_{\mf{g}_1}(P^\mu),
\EE 
where $U(\mf{g}_1)$ is a Hilbert Space representation of an element $\mf{g}_1\in\Pg$ and $G_{\mf{g}_1}$ is a realization of the same element as transformation in the energy-momentum space. 

From this, we formulate the additional condition that assures positivity of energy for at least some irreducible representations of $\Pg$ $^{\underline{8}}$ \footnotetext[8]{ The measurement in such representations yields expected values for four-momentum $\omega(P^\mu)=\Bra{\Psi_\omega}P^\mu\Ket{\Psi_\omega}$ belonging to the linear convex closure of the set $\m{C}_\lambda$, that we denote $\overline{\m{C}_\lambda}_c$,  and is the set of all linear combinations of the type $\lambda p_1^\mu+(1-\lambda)p_2^\mu$, $p^\mu_{1,2}\in \m{C}_\lambda$, $\lambda\in[0,1]$. Being a regular representation, that is, a representation in which $e^{-iX_i t}\Ket{\Psi}$ is continuous for every state $\Ket{\Psi}$,  $\omega_t(P^\mu)=\omega(e^{iXt}P^\mu e^{-iXt})$, it is a continuous path in $\overline{\m{C}_\lambda}_c$ that only achieves the region with $E\leq 0$ by the point $p^\mu=0$, or, equivalently by the state $\omega_{t_0}(P^\mu)=0$, but $\f{d}{dt}\omega_t(P^\mu)\big|_{t=t_0}=\omega([X_{t_0},P^\mu])=\omega(H(P^\mu))=0$, $X_{t_0}=e^{iXt_0}Xe^{-iXt_0}$, where we have made use of (\ref{IdealP}).}:
\BE\label{IdealP}
\begin{split}
[P^\mu,X^\nu]&=c^{(\mu\nu)}_0+c^{(\mu\nu)}_\Ga P^\Ga+c^{(\mu\nu)}_{\Ga\Gb}P^\Ga P^\Gb+\cdots,\\
 &\mbox{ where }c^{(\mu\nu)}_0=0
\end{split}
\EE
$c^{(\mu\nu)}$ being complex coeficients.

The next requirement is that $\Pl$ has the Poincaré Lie algebra as the low energy-momentum limit. A way to assure this is as follows: Let ${C^{ij}_k}^{(0)}$ be the structure constants of the Poincaré Lie algebra
 (\ref{HP})-(\ref{LieM}), and  $C^{ij}_k(E,P)$ algebraic functions of the energy and momentum operators such that $C^{ij}_k(0,0)={C^{ij}_k}^{(0)}$ (Since $[E,P]=0$, we can put the Hilbert space in a representation in which $E$ and $P$ are multiplicative operators). We can define $\mf{p}_{\m{N}}$ as:   
\BE\label{DefLie}
[X_i,X_j]=iX^kC^{ij}_k(H,P).
\EE

It is possible to prove that in a irreducible representation satisfying (\ref{RPnP}) the states can be written as:
\BE\label{EMRepP}
\Psi=\sum_{\sigma}\int d^3\mu(p)\phi(p^\mu,\sigma)\Psi_{p^\mu,\sigma},
\EE
(The conclusion that follows from (\ref{RPnP}) is that the dimension of eigenspace of $P^\mu$ with eigenvalue $p^\mu$ is independent of $p^\mu$)  
 
We write then a state in the form (\ref{EMRepP}) as: 
\begin{gather} 
\Psi_{p^\mu_{\max}}=\int d\mu(p)\phi(p,\sigma)\Psi_{p,\sigma},\nonumber\\
 \mbox{supp}\{ \phi(p,\sigma)\}\in\{E<E_{\max},P^{j}<p^{j}_{\max}\}\label{lowEnergyLim}
\end{gather}
The action of generators of $\Pl$ on those states satisfy:
 \begin{align}
 [X^i,X^j]\act\Psi_{p^\mu_{\max}}&= iX^kC^{ij}_k(H,P)\act\Psi_{p^\mu_{\max}} \nonumber\\
                                 &\to i X^k{C^{ij}_k}^{(0)}\act\Psi_{p^\mu_{\max}}\label{LowLimEP}
 \end{align}
when  $p^\mu_{\max}\to 0$ and $E_{\max}\to 0$. $\mbox{supp}$ denotes the smallest closed set outside which $\phi(p)$ is zero. $\act$ denotes the action (as an operator) in the one particle states, while $\to$ denotes the strong convergence (i.e. $||X^k C^{ij}_k(H,P)\act\Psi-  X^k {C^{ij}_k}^{0}\act\Psi||\to0$)$^{\underline{9}}$ \footnotetext[9]{For selfadjoint operators, or $||X^kC^{ij}_k(H,P)\act\Psi-  X^k{C^{ij}_k}^{0}\act\Psi||\to 0$, or it is not a convergent sequence}. In other words, the typical eigenvalues of the generators effectively change the commutation relations. The property  (\ref{DefLie}) further assures that the algebra generated by $E,P$ is an ideal.

Given the property (\ref{RPnP}), it makes sense to talk about the little group, the subgroup $\m{W}_{p^\mu}$ of $\Pg$, in representation (\ref{RPnP}), that leaves a particular $p^\mu$ invariant. The little group of any $p^\mu$ in the same $\m{C}_\lambda$ is the same (isomorphic). Indeed, in a irreducible representation, any $p^\mu$ can be sent into any other by the action of some $G_{\mf{g}_1}$. Therefore, given $w_1\in \m{W}_{p_1^\mu}$ we can obtain $w_2\in \m{W}_{p_2^\mu}$ by  $w_2=G_{\mf{g}_1}w_1G_{\mf{g}_1}^{-1}$, where $G_{\mf{g}_1}(p_1^\mu)=p_2^\mu$. It follows that, the little group that acts in low energy and momentum should be the same as that at high energies. What leads us to the requirement that every little group of $\Pg$, $\m{W}_{p^\mu}$, for $\lambda\in M_{\Pg}$ must be isomorphic to some little group $\m{W}(\Lambda,p^\mu)$ of the Poincaré group associated to some positive-definite energy irreducible representation. What additionally assures  that the number of internal degrees of freedom for particles is the same as that of relativistic particles, which affects the calculation of the partition function, as in (\ref{LGM}). 
  
A further condition is that $\Pl$ must preserve the $SO(3)$ Lie algebra, the rotation group, and $[H,M_i]=0$, where $M_i$ is the generator of the rotation subgroup along the $i$ direction.  Indeed, it is related to the thermodynamic description of a perfect fluid that is used in non-commutative inflation. That description is no longer valid if the thermodynamic state is no longer rotationally invariant, since the perfect fluid is the one that is isotropic in the reference frame that follows the fluid. In the thermodynamic description, the matter is described by the non-pure state $\Bra{\Psi_T}A\Ket{\Psi_T}=\f{Tr(e^{-\beta H}A)}{Tr(e^{-\beta H})}$, where $A$ is the selfadjoint operator related to an observable and $\beta=\f{1}{k_B T}$. The state is rotation invariant if $\Bra{\Psi_T}RAR^{-1}\Ket{\Psi_T}=\Bra{\Psi_T}A\Ket{\Psi_T}$, where $R$ is a rotation. Since the trace is cyclic, we have: 
\BE
\f{Tr(e^{-\beta H}A)}{Tr(e^{-\beta H})}=\f{Tr(R^{-1}e^{-\beta H}RA)}{Tr(e^{-\beta H})},\forall A \Leftrightarrow [M_i,H]=0
\EE    

\subsection{Conditions on the quantum representation} 

\C{Other thing we usually take for granted is that if some representation of a group has a well defined algebraic structure for generators, all other representations will have too; besides that, they will have the same one. Nonetheless, it is not generally valid. All of these are valid for Lie groups, but if we are no longer dealing with Lie groups, it will be no longer generally valid, the algebraic structure of the generators will no longer be a concept attached to the group itself but to specific representations.} 

It leads us to the last condition, which is related to the application of prescription 1. (\ref{FockHamiltonian}) must be a generator of the $\Pl$ if $H\in \Pl$. Although this is a subtle point, it is the main point of this paper. (\ref{FockDef}) is actually a valid representation of the group $\Pg$, since it satisfies (\ref{R1})-(\ref{R3}), but, at least that it is a Lie Group, the generators of the one-parameter subgroups are not assured to form the same algebra as $\Pl$. If it is the case, it will mean that we do not have a uniquely defined notion of a dispersion relation that replaces the relativistic one. In particular, the generators of these subgroups may not form an algebra at all, not satisfying the above discussed requirements. If the direct product of two one-particle representations do form an algebra satisfying all the above requirements, but do not form the same algebra than that of the one particle subspace, then an $S$ matrix connecting a one-particle with the two particles would not be compatible with energy conservation in all reference frames, since the energy of in and out states would not transform as the same function as in Eq.(\ref{RPnP}).
 
This leads us to consider two kinds of representations of the group $\Pg$, the one whose generators of one-parameters subgroups do not form the same algebra than $\Pl$, that we will call the $\pi_{\mf{g}}$ representation, and that one whose generators of the one-parameter subgroups do form the same algebra as $\Pl$, that we will call the $\pi_{\m{A}}$ representation.

\section{The non-existence of a Lie group that realizes non-commutative inflation}

In our previous work \cite{Machado}, we  obtained a set of conditions to be satisfied by a dispersion relation that assures minimal desirable conditions for inflation, like the minimum duration of it, related to problems like the flatness of the universe (\cite{BrandenbergerRevInf2} and \cite{BrandenbergerRevInf}). These conditions were given by a set of inequalities that contains the particular ansatz applied by the original authors of non-commutative inflation \cite{Noncommutative_Inflation}. It was shown that, given the original equations of the model, (\ref{rho})-(\ref{rhoT}), the minimum duration of inflation is related to a limitation of the photon's momentum (a consequence of condition 3 of the theorem 2 in \cite{Machado}).

Since Casimir is an algebraic entity of $\Pl$, those conditions should be included in the set of algebraic conditions for inflation discussed in the last section. Let us show here that the limitation of the photon's momentum is precisely what implies that (\ref{FockHamiltonian}) cannot be the time translation generator of a representation of $\Pl$, provided that all $C_\lambda$ (the mass shell deformation) with positive definite energy is connected and allows a zero momentum state, that is, $C_\lambda$ contains the point $(E,P)=(E^{(0)}_\lambda,0)$.    

Consider the deformed forward light cone $V^+_{\Pl}$ defined in the previous section. Suppose that $C(E,P)=\lambda_P\in M_{\Pl}$ ($M_{\Pl}$ defined in the previous section) for the irreducible representation that describes radiation. $\m{C}_{\lambda_P}$ should contain the point $(E,P)=(0,0)$ since, in the low momentum and energy limit, the deformed dispersion relation should be the usual photon's dispersion relation. Consider additionally that for any other $\lambda\in M_{\Pl}$,  $\m{C}_\lambda$ contains the point $(E,P)=(E^{(0)}_{\lambda}=m^2,0)$, i.e. we are assuming a relativistic low energy-momentum limit for every particle's deformed dispersion relation associated to $\Pl$.  Since $\m{C}_{\lambda}$ is rotationally invariant by hypothesis (a condition related to the perfect fluid description),  let us consider the graph relating the momentum's magnitude to the energy in a particular irreducible representation with the dispersion relation Casimir $\lambda$: $||P||=P_{\lambda}(E)$.

 It happens that $P_{\lambda}(E)\geq 0$ and there is no $E$ such that $P_{\lambda_1}(E)=P_{\lambda_2}(E)$ for $\lambda_1\neq \lambda_2$, since $C(E,P)$ is an analytic function. Therefore   $P_{\lambda}(E^{(0)}_\lambda)=0< P_{\lambda_P}(E_\lambda)$, which implies that $P_{\lambda}(E)<P_{\lambda_P}(E)$ for all $E$. Indeed, if there is an $E$ such that $P_{\lambda}(E)>P_{\lambda_P}(E)$, it follows by the continuity of $P_{\lambda}(E)$ (by analyticity of the Casimir $C(E,P)$) that there must be an $E$ such that $P_{\lambda}(E)=P_{\lambda_P}(E)$. But such a point cannot exist. The conclusion is that if each $\m{C}_\lambda$ with positive definite energy is connected (to use the continuity argument above) and has relativistic energy in the low energy-momentum limit, a limitation of photon momentum implies a limitation of momentum for all positive definite energy irreducible representations.

The reasoning above is only valid for irreducible representations. Let us now argue about the general case. To do that, let us consider an important result in the representations of Von Neumann algebras$^{\underline{10}}$ \footnotetext[10]{\label{VNA}A Von Neumann algebra $\m{A}$ is a subalgebra of the operators in the Hilbert space $\m{B}(\m{H})$ that is an $*$-algebra, that is, it contains the adjoint of each element, and is equal to its bicommutant, that is, $\m{A}=\m{A}''$. The bicommutant of $\m{A}$ is the set of all operators in the Hilbert space that commutes with the commutant $\m{A}'$, where the commutant is the set of all operafors that comute with $\m{A}$. } that additionally is going to be the basis of a possible new procedure to obtain the correct non-relativistic inflationary equation of state. This result comes from a powerful form of the spectral theorem. Indeed, in quantum mechanics, we know that, given an irreducible representation of the Heisenberg algebra, there exists a unitary transform $U$ such that we can put the position or momentum operator (not at same time) as multiplicative operators in the Hilbert space $L^2(\mathbb{R})$ of the square (Lebesgue) integrable functions, that is: $\hat{x}\Ket{\Psi}=x\Psi(x)$, $\Psi(x)\in L^2(\mathbb{R})$. This diagonal form, however, is not general. It is only valid for the so called cyclic operators$^{\underline{11}}$\footnotetext[11]{A selfadjoint operator $A$ in some domain $\m{D}$ in the Hilbert space $\m{H}$ is cyclic if there exists some $\Psi\in \m{D}$ such that the set $A^n\Psi$, for all $n$, spans the entire Hilbert space}  with spectrum equal to $\mathbb{R}$.  The general form of this decomposition is the so called direct integral representation (see \cite{Gelfand} for a proof).

Given a self adjoint operator $A$ in some domain $\m{D}$ in a separable (i.e., with countable basis) Hilbert space $H$, there exists a unitary transform $U:H\to \mf{h}$ such that
\BE
\mf{h}=\int^\oplus \m{H}(\lambda) d\mu(\lambda), 
\EE
$\m{H}(\lambda)$ a (possibly infinite dimensional) separable Hilbert space for each $\lambda$. This is the Hilbert space whose vectors are written as 
\BE
\Psi=\int^ \oplus \Psi(\lambda) d\mu(\lambda), \quad \Psi(\lambda) \in \m{H}(\lambda) 
\EE 
and the inner product is:
\BE
(\Phi,\Psi)=\int \Le<\Phi(\lambda),\Psi(\lambda)\Ri>_{\lambda}d\mu(\lambda), 
\EE
$\Le<\Phi(\lambda),\Psi(\lambda)\Ri>_{\lambda}$ being the inner product of the Hilbert space $\m{H}(\lambda)$, $\mu(\lambda)$ a suitable integration measure$^{\underline{12}}$\footnotetext[12]{There is an additional detail that there exists a countable set of mensurable sets $X_i$ of $\lambda$ values such that the dimension of the Hilbert space  $\m{H}(\lambda)$ is the same for the same $X_i$ and, when it is an infinite dimensional Hilbert space for some $X_i$, it is a separable Hilbert space.}. Furthermore, we have: 
\BE
UAU^{-1} \int^ \oplus \Psi(\lambda) d\mu(\lambda)=\int^ \oplus \lambda\Psi(\lambda) d\mu(\lambda).
\EE
The domain of the operator  $UAU^{-1}$, $U(\m{D})$, is the set:
\BE
\int ||\lambda\Psi(\lambda)||_\lambda^2 d\mu(\lambda)<\infty,
\EE
$||*||_\lambda$ being the norm of the Hilbert space $\m{H}(\lambda)$.

We can do the same thing with a set of commuting self-adjoint operators $A_i$ with a common domain $\m{D}$ $^{\underline{13}}$\footnotetext[13]{Actually, $A_i$ is required to be essentially self-adjoint in the domain $\m{D}$}, defining the unitary transform (this is actually the Von Neumann's theorem for Abelian Von Neumann algebras):
\BE\label{MultiU}
U_{\m{D}}:\m{H}\to\mf{h}=\int^\oplus \m{H}(\lambda_1,\lambda_2,\cdots) d\mu(\lambda_1,\lambda_2,\cdots),
\EE
such that $U A_i U^{-1}\int^\oplus \psi(\lambda_1,\cdots \lambda_N)d\mu(\lambda_1,\cdots \lambda_N)= \int^\oplus \lambda_i \psi(\lambda_1,\cdots \lambda_N)d\mu(\lambda_1,\cdots \lambda_N)$. 

We can define the operator $X$ in $\mf{h}$ by defining the operator $X(\lambda)$ in some domain $\m{D}(\lambda)$ in each $\m{H}(\lambda)$:
\BE
X\Psi=\int^{\oplus} X(\lambda)\Psi(\lambda)d\mu(\lambda).
\EE
We denote:
\BE\label{OpX}
X=\int^{\oplus} X(\lambda)d\mu(\lambda)
\EE
as the operator whose domain is the set $\int^\oplus \Psi(\lambda) d\mu(\lambda)$, $\Psi(\lambda)\in\m{D}(\lambda)$ and $\int ||X(\lambda)\Psi(\lambda)||^2_\lambda d\mu(\lambda)<\infty$.  
It is possible to prove that the algebraic function $f[X]$ of $X$ can be writen as:
\BE
f[X]=\int^{\oplus} f[X(\lambda)]d\mu(\lambda),
\EE 
which shows the algebraic independence of the operators defined in each $\m{H}(\lambda)$. 
 
Now suppose a representation of $\Pl$ in the Hilbert space $H$ and $C_1,C_2,\cdots, C_N$ is the set of Casimir operators of the representation. By the uniform transform we put $H$ in a direct integral form such that all Casimir operators are multiplicative:
\BE\label{DirIntPl}
\mf{h}=\int^\oplus \m{H}(\lambda) d\mu(\lambda),
\EE
where $\lambda$ is the entire set of Casimir eigenvalues. But in each $\m{H}(\lambda)$, the Casimir is a multiple of the identity, therefore, by Schur's lemma, there must exist an irreducible representation of $\Pl$ on each $\m{H}(\lambda)$ whose generators are written in the form (\ref{OpX}). 

We observe that the set of Casimir operators can include operators that are not predicted by the algebra of $\Pl$, that is, other than (\ref{W}), in the case of Poincaré Lie algebra. Indeed, suppose $\pi:\Pl \to \m{O}(\m{D})$  ($\m{O}(\m{D})$ stands for operators in the domain $\m{D}$) is an irreducible representation, and the algebra predicts a particular Casimir set $\m{C}_1,\m{C}_2,\cdots$ written as algebraic functions of the generators of $\Pl$. We can write another representation in terms of the irreducible one: $\pi'=\pi\oplus\pi$. This is a reducible representation, but there is no other Casimir in the image of the representation $\pi'$ other than $\pi'(\m{C}_i)$. However, $C''=(a \pi[\m{C}_i])\oplus (b \pi[\m{C}_i])$, $a,b\in \mathbb{R}$, $a\neq b$, is a Casimir operator that is not a multiple of the identity and it is not in the image of the original representation. This is because the representation of an algebra like the Lie algebra (actually its enveloping algebra) is not a Von Neumann algebra (see footnote (\ref{VNA})) such that $\m{A}=\m{A}''$, that is, that contains all its Casimir elements. It is contained in some Von Neumann algebra.  Those additional Casimirs are related to the degenerance of the representation, i.e., the number of times, finite or infinite, that it appears in the decomposition.

We then represent this decomposition as:
\BE\label{DirIntSigma}
\mf{h}=\int^\oplus \m{H}(\lambda,\sigma) d\mu(\lambda,\sigma),
\EE
where $\lambda$ is the algebraic Casimir eigenvalues, and $\sigma$ is related to the desgenerance. 

Now suppose that in the decomposition of the representation of $\Pl$ there appears irreducible representations $\pi_\lambda$ with dispersion relation eigenvalues $\lambda_1\notin M_{\Pl}$ for some set $B$ of $\lambda$ with non-zero measure. Suppose also there is a mensurable function $f(\lambda)$ such that $\int d\mu(\lambda) |f(\lambda)|^2 < \infty$.  If $B$ has a finite measure, this function can be chosen as the characteristic function of $B$, $\chi_B$, which is equal to unity in $B$ and zero otherwise. We can choose a normalized vector $\psi_{\lambda}$ of negative energy $E_{\lambda}$ in each $\m{H}(\lambda)$ with $\lambda \in B$ and the null vector in every other $\lambda$ value. The state $\int^\oplus d\mu(\lambda) f(\lambda)\psi_{\lambda}$ belongs to the Hilbert space  $\mf{h}$ and has negative energy.  The conclusion is that for all direct integral decompositions of $\Pl$ with positive definite energy, only irreducible representations with $\lambda_0\in M_{\Pl}$ do appear (except by zero measure sets).

Furthermore, we can write $P^\mu$ as a multiplicative operator if we apply a further unitary transform in  $\mf{h}$ that represents each $\m{H}(\lambda,\sigma)$ in (\ref{DirIntSigma}) as the Hilbert space of $N$ component functions$^{\underline{14}}$ \footnotetext[14]{As stated, this decomposition is valid for irreducible representations if the associated Lie algebra has space-time translation generators that form an ideal} defined in the place of the points  $(E,\vec{P})\in \mathbb{R}^4$ in the energy-momentum space that satisfies $C(E,P)=\lambda_1$  with measure $\nu$$^{\underline{15}}$ \footnotetext[15]{That is, the inner product of the irreducible representation is $(\Phi,\Psi)=\int_{\m{C}_\lambda} \psi^*(p^ \mu)\psi (p^ \mu)d\nu=\int_{\m{C}_\lambda} \sum_\sigma \psi_\sigma^*(p^ \mu)\psi_\sigma (p^ \mu)d\nu$, where $\psi_\sigma$ are the components of $\psi$. } ( i.e. $(E,P)\in\m{C}_{\lambda_1}$, recalling that $\lambda=(\lambda_1,\lambda_2,\cdots)$). Combining the measures in $\m{C}_{\lambda_1}$ and $\lambda_1$ $^{\underline{16}}$ \footnotetext[16]{If we have two measurable sets $X_1$ and $X_2$ with measures $\nu_1$ and $\nu_2$, we can define a new measurable set $X_1\times X_2$ in which all measurable sets are of the Cartesian product form $A_1\times A_2$ for $A_i$ a measurable set of $X_i$ and the measure of $A_1\times A_2$ is $\nu_1(A_1)\cdot \nu_2(A_2)$. The measure of (\ref{DirIntPl}) is the Cartesian product form}, we define a new measure $\mu'$ in four-momentum space and write:      
\BE\label{DMSID}
\mf{h}=\int^\oplus \m{H}(p^\mu,\lambda',\sigma) d\mu'(p^\mu)d\nu(\lambda',\sigma),
\EE
where $\lambda'$ denotes the remaing algebraic Casimir elements $\lambda'=(\lambda_2,\cdots)$. The support of the measure $\mu'$, i.e., the region giving contributions to the integral$^{\underline{17}}$ \footnotetext[17]{the largest closed set in which every open neighborhood has positive measure}, is the set of points belonging to $\m{C}_\lambda$ for each $\lambda$ belonging to the direct integral decomposition. For simplicity of notation, let us denote
\BE
\mf{h}=\int^\oplus \m{H}(p^\mu,\tau) d\chi,
\EE
where $\tau=(\lambda',\sigma)$.

We have the operator $P^\mu$ defined in $\mf{h}$ by the operator ${P^\mu}_{(p^\mu,\tau)}$ in each $\m{H}(p^\mu,\tau)$ such that, for all $\Psi_{(p^\mu,\tau)}\in \m{H}(p^\mu,\tau) $: 
\BE
P_{(p^\mu,\sigma)}^\mu{\Psi_{(p^\mu,\sigma)}}=p^\mu{\Psi_{(p^\mu,\sigma)}},\quad p^\mu\in V^+_{\Pl} \subseteq \mathbb{R}^4.
\EE 
Then, for all normalized ${\Psi}\in \mf{h}$, $\Bra{\Psi}P^\mu\Ket{\Psi}\in \overline{V^+_{\Pl}}_c$, where $\overline{V^+_{\Pl}}_c$ denotes the convex linear closure $^{\underline{18}}$ \footnotetext[18]{The set of all points written in the form  $\Ga p^\mu_1+(1-\Ga) p^\mu_2$, $\Ga\in[0,1]$ e $p^\mu_1,p^\mu_2\in V^+_{\Pl}$. $\overline{V^+_{\Pl}}_c$ is obtained joining to $V^+_{\Pl}$ every line segment that joints any two points of $V^+_{\Pl}$.}. 

Indeed, we can partition the set $V^+_{\Pl}$ in regions $U_n=V^+_{\Pl}\cap R_{ijkl} $, where $n$ is a set of indexes $n=(i,j,k,n)$ and $R_{ijkl}$ are the rectangles:  
\BE
R_{ijkl}=[p^0_i,p^0_{i+1})\times [p^1_j,p^1_{j+1})\times [p^2_k,p^2_{k+1})\times [p^3_l,p^3_{l+1}),
\EE 
obtained from partitions  $p^\mu_i$, $i=1,2,\cdots N$, $\mu=0,1,2,3$, such that $\bigcup_{n}R_{n}\supset V$. 

It follows that $\Bra{\Psi}P^\mu\Ket{\Psi}=\int d\chi p^\mu||\psi(p^\mu,\tau)||^2$, which implies the inequalities:   
\begin{gather}
 \sum_n \inf_{U_n}(p^\mu) \int_{U_n}||\psi(p^\mu,\tau)||^2 d\chi \leq \Bra{\Psi}P^\mu\Ket{\Psi}\\
 \Bra{\Psi}P^\mu\Ket{\Psi}\leq \sum_n \sup_{U_n}(p^\mu) \int_{U_n} ||\psi(p^\mu,\tau)||^2 d\chi, 
\end{gather} 
but $||\psi(p^\mu,\tau)||^2$ is positive definite and $\int_{V^+_{\Pl}} d\chi ||\psi(p^\mu,\tau)||^2=1$, in such a way that the right side of the first inequality and the left side of the last one  can be written as: 
\BE 
\sum_{n} p_n^\mu c_n,\mbox{where } \sum_n c_n=1\mbox{, } p^\mu_n\in V^+_{\Pl}\mbox{ and } c_n\geq 0.
\EE 

Suppose that $\sum_{n=0}^N p_n^\mu c_n\in \overline{V^+_{\Pl}}_c $ for $\sum_{n=0}^N c_n=1$, then $\sum_{n=0}^{N+1} p_n^\mu c'_n\in \overline{V^+_{\Pl}}_c$ for $\sum_{n=0}^{N+1} c'_n=1$. Indeed, we can write 
\BE 
\sum_{n=0}^{N+1} p_n^\mu c'_n=\f{\sum_{n=0}^{N} p_n^\mu c'_n}{\sum_{n=0}^N c'_n}\f{\sum_{n=0}^N c'_n}{\sum_{n=0}^{N+1} c'_n}+p^\mu_{N+1}(1-\f{\sum_{n=0}^N c'_n}{\sum_{n=0}^{N+1} c'_n})
\EE

But for $N=1$ we have $p^\mu_0c_0+p^\mu_1(1-c_0)\in \overline{V^+_{\Pl}}_c $.

Therefore, if $p^\mu\in V^+_{\Pl}$ for every $(p^\mu,\sigma)$, $||\vec{p}||<p_{\max}$, it follows that for every representation of $\Pl$ with positive definite energy the moment is limited by photon's momentum: 
\BE
\Bra{\Psi}{P^j}\Ket{\Psi}\leq p^j_{\max},
\EE
since every convex linear combination with limited momentum has the same bound to the momentum.

But the generators associated to representation (\ref{FockDef}) are of the form: 
\BE
X_F=\sum_{N=1}^{\infty} { }^{\oplus} (X\otimes I \cdots \otimes I+ I\otimes X\otimes \cdots \otimes I+\cdots+ I\otimes I\cdots \otimes X)_N
\EE

In particular, the energy of this representation is positive definite and the moment is unbounded (it assumes every integer multiple of the irreducible representation momentum). Therefore, the generators of (\ref{FockDef}) do not realize $\Pl$. Besides, since every representation of a Lie group induces a representation of the same Lie algebra and (\ref{FockDef}) is a representation of the group $\Pg$, there is no Lie group that realizes non-commutative inflation.  

\section{The alternative prescription for the non-relativistic inflationary equation of state}

The conclusion of the previous section is that, given the condition for minimum duration of inflation in non-commutative inflation \cite{Machado}, which implies that the momentum of an individual photon is limited, and assuming that all $C_\lambda$ (the mass shell deformation) with positive definite energy is connected and allows a zero momentum state, the non-commutative inflation is then necessarily associated with a non-trivial Hopf algebra structure with a non-trivial coproduct structure. Therefore Eq. (\ref{FockDef}), which leads to the fundamental equations of non-commutative inflation, cannot be the appropriate time-translation generator, according to the discussion of section (\ref{CAlS}). Considering the group point of view, that is, considering that we are representing the group $\Pg$ whose generators are $e^{-iX_it}$ for $X_i\in\Pl$,  to define the physics, we are restricting the possible representations that are used in such a construction, excluding representation (\ref{FockDef}). This kind of restriction is not new in physics, since we already restrict the representations of the Poincaré group that can appear in a relativistic quantum theory to those of positive definite energy, for example.   

If we are able to find a Hopf algebra deformation $\Pl$ satisfying all the requirements of  section (\ref{CAlS}) and, additionally, a coproduct  $\Delta^{\Pl}$ (defined in section (\ref{Hopf}))  satisfying:
\begin{multline}
\Le(\Psi_{p^\mu_{\max}}\otimes\Phi_{p^\mu_{\max}},\Delta^{\Pl} X^{\Pl}_i \Psi_{p^\mu_{\max}}\otimes\Phi_{p^\mu_{\max}}\Ri)\\ \to \Le(\Psi_{p^\mu_{\max}}\otimes \Phi_{p^\mu_{\max}}, \Delta^{\mf{p}} X^{\Pl}_i\Psi_{p^\mu_{\max}}\otimes \Phi_{p^\mu_{\max}}\Ri),
\end{multline} 
where $\Psi_{p^\mu_{\max}}$ and $\Phi_{p^\mu_{\max}}$ are normalized states written as in Eq. (\ref{lowEnergyLim}) and the limit is when $(E,P)\to (0,0)$. $X^{\Pl}_i $ is the Hopf algebra deformation of the generator $X^{\mf{p}}_i$ of the Poincaré Lie algebra and $\Delta^{\mf{p}}$ is the Lie algebra coproduct:  $\Delta^{\mf{p}}X=X\otimes I+I\otimes X$. This rule means that the new coproduct  $\Delta^{\Pl}$ becomes the usual coproduct of Lie algebras for low energy and momentum. 

Given such a Hopf algebra deformation, with the appropriate coproduct, we can define the time translation generator that replaces Eq. (\ref{FockDef}) and is part of a representation of $\Pl$:
\BE\label{HopfH}
H_F^{\Pl}={\sum_{N=0}^\infty}^{\oplus}{\Delta^{\Pl}}^N H,
\EE
where ${\Delta^{\Pl}}^N$ is defined in Eq.(\ref{DeltaN}) for generic coproduct rule $\Delta$. In some sense, it implies that particles interact even in the free theory, that is, they do not evolve in time in a completely independent way. 

That prescription has the disadvantage that we can only apply it to deform the free fields. Besides that, we need a full definition of the algebra and Hopf algebra structure associated to a particular photon phenomenological dispersion relation, which includes, in particular, the assurance of the existence of a Hopf algebra associated to it. 

It is however easy to assure the existence of a algebra (without considering its coalgebraic structure, i.e., its coproduct, counity and antipode) that satisfies all the requirements discussed in section (\ref{CAlS}) and leads to a particular photon dispersion relation$^{\underline{19}}$ \footnotetext[19] {In fact, consider $X_i$ the infinitesimal generators of the scalar representation of the homogeneous Lorentz group in momentum space:
\BE
\phi(x^\mu)\to\phi(\Lambda x^\mu)
\EE 
They are contravariant vectors, or first order differential operators. Consider the diffeomorphism $\Phi:(E,p)\to(\bar{E},\bar{p})$ given by $\bar{E}=E$ ; $\bar{p}=p/f(E)$, the $f(E)$ given in (\ref{FDR}). Define the new algebra as multiplication operators $E$, $p$ and $\bar{X}_i=\Phi_{*}X_i$, the pushfoward operator $\Phi_{*}X_i\act f=X_i \act f\circ \Phi$. This new algebra satisfy $[\bar{E},\bar{p}]=0$; $[\bar{X}_i,\bar{X}_j]=C^k_{ij}\bar{X}_k$, $C^k_{ij}$ the same structure constants of the original Lorentz group,  but $[\bar{X}_i,E]=(\bar{X}_i\act E)=F(E,p)$ and $[\bar{X}_i,p]=(\bar{X}_i\act p)=G(E,p)$; As required $[C(E,p),E]=[C(E,p),p]=[C(E,p),\bar{X}_i]=0$, where $C(E,p)=f^2(E)p^2-E^2$. All the conditions of the section (\ref{CAlS}) are satisfied}.  We can use an alternative prescription only based on algebra properties.  This can be applied to interacting field theory as well as to free fields and assuring that we are constructing a physics that is based on representations of a group $\Pg$ that realizes the same algebra of generators $\Pl$. That is, based on $\pi_{\m{A}}$ representations.   

The basic idea is exploring the fact that every relativistic quantum field theory, associated to some Lagrangian density, has associated with it a representation of the Poincaré Lie group in the positive definite Hilbert space (i.e. such that the norm of states is positive). This representation enters in the relativistic covariance law of the fields:
\BE\label{CovarianceLaw}
{A^i}_j[(\Lambda,a)]\phi_i(\Lambda x+a)= U[(\Lambda,a)]\phi_j(x) U^{-1}[(\Lambda,a)] 
\EE
where ${A^i}_j[(\Lambda,a)]$ is a finite dimensional representation of the Poincaré Lie group. It happens that (\ref{CovarianceLaw}) may not be satisfied in the positive definite Hilbert space, as in the case of electrodynamics, but we can restrict the action of $U[(\Lambda,a)]$ to the equivalence class of  physical states (something similar to what we have done in the GNS construction to construc equivalence classes that describe the same state).       

This representation can be recovered from the Wightman functions by the so called Wightman reconstruction theorem, that is an application of the GNS construction. Alternatively, we can obtain it by other classes of functions like the time-ordered Green's functions or the Schwinger functions (see some original papers on this subject \cite{Eckmann}, \cite{OsterwalderI} and \cite{OsterwalderII}). In the case of free fields, we already have that representation, it is simply (\ref{Fock}) for a suitable choice of irreducible representation of Poincaré group.

We then put the associated representation in the direct integral form:
\BE
\mathcal{U}^{\mathcal{P}}=\mathcal{U}_0\oplus\int^{\oplus} d\mu(\lambda,\sigma)\mathcal{U}_{\lambda,\sigma}^{\mathcal{P}},
\EE
that is defined in:
\BE
\mathcal{H}^{\mathcal{P}}={c\Psi_0}\oplus\int^{\oplus} d\mu(\lambda,\sigma)\mathcal{H}_{\lambda,\sigma}^{\mathcal{P}},
\EE
where $\mathcal{U}_0$ is the trivial vacuum representation, that leaves the vacuum subspace $c\Psi_0$ invariant, while $\lambda$ denotes the eigenvalues of those Casimir elements predicted by the Poincaré Lie algebra (\ref{W}) and $\sigma$ denotes the Casimir eingenvalues related to degenerance. That is, the irreducible representation $\mathcal{U}_{\lambda,\sigma}^{\mathcal{P}}$ is the same for all $\sigma$ and we have a $\sigma$ value for each time that the same representation appears in the decomposition.

We then postulate:   
\begin{Pres}[$\pi_{\m{A}}$]
Be $\Pl$ an algebra (or Hopf algebra) with generators $X_i$ and $\Pg$ the group generated by one parameter subgroups of the form $e^{-iX_i t_i}$, $t_i\in \mathbb{R}$. Be $\pi_{\lambda'}:\Pl\to O(\m{D})$ the irreducible representation of $\Pl$ in Hilbert space (obtained by GNS construction with states in $\Pl$ or $\Pl^*$) associated to algebraic Casimir eingenvalues $\lambda'=(\lambda_1,\lambda_i,\cdots)$. Be $\mathcal{U}_{\lambda'}^{\mathcal{P}_{\m{N}}}$ the representation of $\Pg$ generated by $e^{-i\pi_{\lambda'}(X_i) t_i}$. The representation of $\Pg$ that deforms a relativistic quantum field theory is obtained by the associated Poincaré Lie group representation (restricted to positive definite Hilbert space of physical states) put in the direct integral form:
\BE
\mathcal{U}^{\mathcal{P}}=\mathcal{U}_0\oplus\int^{\oplus} d\mu(\lambda,\sigma)\mathcal{U}_{\lambda,\sigma}^{\mathcal{P}},
\EE
by the replacement:
\BE\label{GenHopfPres}
\m{U}^{\m{P}}_{\lambda}\to \mathcal{U}_{\lambda'}^{\mathcal{P}_{\m{N}}}, 
\EE
for the set of Casimir eigenvalues $\lambda'$ satisfying the condition:
\begin{multline}\label{ConvCon}
\left(\mf{u}(\Psi^{\lambda}_{p^\mu_{\max}}),e^{-i\pi_{\lambda'}[X^{\Pl}_i] t_i}\mf{u}(\Phi^{\lambda}_{p^\mu_{\max}})\right)\\ \to\left(\Psi^{\lambda}_{p^\mu_{\max}},e^{-i\pi_{\lambda}[X^{\mf{p}}_i] t_i} \Phi^{\lambda}_{p^\mu_{\max}}\right).
\end{multline}
when $(E_{max}, p^j_{max})\to (m,0)$ for normalized states $\Psi^{\lambda}_{p^\mu_{\max}}$ and $\Phi^{\lambda}_{p^\mu_{\max}}$ written in the form (\ref{lowEnergyLim}).
\end{Pres}
Here, $\mf{u}:\m{H}_1\to\m{H}_2$ defines an unitary transform connecting the Hilbert spaces where the representations of the Poncaré group and $\Pg$ are defined. As before, $X^{\Pl}$ is the corresponding operator in $\Pl$ to the Poincaré Lie algebra element $X^{\mf{p}}$. $m$ is the relativistic energy at zero momentum that, by hypothesis, is the same for the corresponding $\Pg$ representation. 

This prescription defines the new time translation generator in the direct integral form:
\BE\label{NewH}
{H}^{\Pl}={0}\oplus\int^{\oplus} d\mu(\lambda,\sigma){H}_{\lambda',\sigma}^{\Pl},
\EE
where ${H}_{\lambda',\sigma}^{\Pl}$ is the Hamiltonian of the irreducible representation of $\Pl$ associated to the set of eigenvalues $\lambda'$ ( that is diagonal in the energy-momentum representation and related to the momentum by $C({H}_{\lambda,\sigma}^{\Pl},P)=\lambda_1$). That is, not only the photon dispersion relation enters in the Hamiltonian of the theory but the dispersion relations of all of the species of particles.

The condition (\ref{ConvCon}) is actually stronger than (\ref{LowLimEP}) and  allows us to assure the convergence of the $\Pg$ representation to the original relativistic representation in the weak sense (i.e., in the sense of matrix elements: $\left(\mf{u}(\Psi),\mathcal{U}^{\mathcal{NP}}\mf{u}(\Phi)\right)\to\left(\Psi,\mathcal{U}^{\mathcal{P}}\Phi\right)$, where $\mf{u}$ is a unitary map that connects the Hilbert spaces where the Poincaré representation is defined  to the one where the $\Pg$ representation is defined) if all intermediate states of direct integral decomposition converge as (\ref{lowEnergyLim}). This corresponds to states of sufficiently low energy and momentum $^{\underline{20}}$ \footnotetext[20]{Lebesgue's theorem says that if $g_N(x)\to g(x)$ for all $x$ and $|g_N(x)|\leq f(x)$ for some integrable $f(x)$, then $g(x)$ is Lebesgue integrable and $\int g_N(x)\to\int g(x)$. Since we know that, for the original QFT, there exists Hilbert space vectors $\Psi_{\sigma,\lambda}$ in each $\mathcal{H}_{\lambda,\sigma}$ such that the inner product integral converges, and the unitary transformation satisfies $\left|\left(\mf{u}(\Psi_{\lambda,\sigma}),\mathcal{U}^{\mathcal{NP}}_{\lambda',\sigma'}\mf{u}(\Phi_{\lambda,\sigma})\right)\right|\leq||\Psi_{\lambda,\sigma}||\cdot||\Phi_{\lambda,\sigma}||\leq||\Phi_{\lambda,\sigma}||^2+||\Psi_{\lambda,\sigma}||^2$ which is a measurable function with finite integral, we have the fulfillment of conditions of the Lebesgue theorem.  }.

\section{Conclusion}

In a previous paper \cite{Machado}, we showed that the fundamental equations of non-commutative inflation, first obtained in \cite{NVSL} and applied in a model of inflation in \cite{Noncommutative_Inflation}, can actually be related to a representation of some group that deforms Poincaré Lie group, a group that can be constructed by starting from a Hopf algebra (allowing a possible connection with the original motivation of non-commutative spaces, along the lines of \cite{Hopf}). We showed that the condition of minimum duration of inflation is actually related to a limitation of the momentum of an individual photon. In the current paper, we showed that a group that replaces Poincaré must satisty some important physical constraints already satisfied by the Poincaré Lie group. These constraints  do not depend on cosmological requirements and affect not only the group's abstract algebraic properties, but the representations themselves. Considering these constraints, the limitation of the photon's momentum, together with the minimal additional conditions on the deformed mass shell, such as being connected at least for those solutions of positive definite energy and the same solutions allowing a zero momentum state, implies that the notion of Fock space is not suitable and a Hamiltonian like (\ref{FockHamiltonian}) cannot be the starting point of the fundamental equations of the non-commutative inflation.

Everything starts by the fact that we are replacing the Poincaré group $\m{P}$ by a group $\Pg$  that has connected one-parameter subgroups that allow us, by Wigner and Stone theorems, to define generators of the representations of these one-parameter subgroups as self-adjoint operators that we assume that form an algebra $\Pl$. \C{ A thing that we usually take for granted is that, not matter what is the group we are talking about, the generators will always for an algebra. Nevertheless, that is not mathematically true. However, as we explained in section \ref{CAlS}, Physics cannot be formulated without an underlying algebraic structure for symmetry generators.  Other thing we usually take for granted is that if some representation of a group has a well defined algebraic structure for generators, all other representations will have too; besides that, they will have the same one. It is not generally valid too. All of these are valid for Lie groups and, once we are no longer dealing with Lie groups, it is no longer generally valid. Once we leave the domain of Lie groups,  the algebraic structure of the generators is not longer a concept attached to the group itself but to specific representations.  We show, however, that if we don't have the very same algebra of generators for all symmetry representations that we apply in physics, physics will not be consistent.}   

 Essentially, we show that, although a Fock space representation (\ref{FockDef}) is a valid mathematical representation for $\Pg$, in the sense that it preserves the group structure given by Eqs.(\ref{R1})-(\ref{R3}), for the kind of deformed dispersion relation that drives inflation with minimum duration, this representation does not satisfy the same algebra of generators used in the irreducible representations. As a consequence, non-commutative inflation is not related to some Lie group, since every representation of a Lie group induces a representation of the very same Lie algebra of generators. 

 As we argue, the use of this Fock like representation is a problem for a physical theory, since it might imply that a stable state in a particular reference frame is an unstable state in another one, or a possible S matrix connecting a one particle state with a $n$-particle state does not conserve energy in all reference frames. In particular, we do not have a unique well defined notion of a dispersion relation that replaces the usual relativistic one.
 
  It does not mean that we cannot use a group like $\Pg$ in physics. It means that we must restrict the class of representations of $\Pg$ that is used to construct physics, something similar to restricting the representations of the Poincaré group that appear in QFT to those with positive definite energy. But now, besides that, we must restrict the representations of the $\Pg$ group to those with positive definite energy and that induces a representation of the same algebra of generators $\Pl$.\C{ We here call these representations $\pi_\m{A}$.}   

An alternative way to say that (\ref{FockDef}) is not a suitable representation of $\Pg$ is to say that (\ref{FockHamiltonian}) is not a time-translation generator for $\Pl$ if $H$ is.  Since, in the Hopf algebra language, (\ref{FockHamiltonian}) is the coproduct rule for Linear Lie algebras, this suggests to us that  non-commutative inflation is actually related to some non-trivial Hopf algebra, by understanding that a Hopf algebra is an algebra with a rule to construct tensor products that satisfy the same algebra. 

The Hopf algebra concept is a possible link between the non-commutative space concept and non-commutative inflation, since an alternative interpretation for Hopf algebras is that it is a set of transformations that acts on other algebraic structures in a way dependent on the product rule. The product rule is the essential information of non-commutative spaces (see \cite{Strocchi}). We can therefore use the coproduct rule of Hopf algebra to define the suitable non-commutative inflationary Hamiltonian that defines the canonical partition function that describes the radiation thermodynamics as in (\ref{HopfH}). This is our first possible prescription to define the non-commutative inflation. This prescription, however, needs the full knowledge of the Hopf algebraic structure associated to some phenomenological dispersion relation, a problem by itself, and can only be applied to free fields. 

We can consider an alternative prescription that is based only on algebraic aspects, that is, it does not depend on the coalgebraic aspects, which means that it does not depend on the information that differs the Hopf algebra from a simple algebra, as the coproduct. This prescription is based on the infinite dimensional generalization of the concept of direct sum, the direct integral.  It was originally proposed by us in \cite{Machado}, as an alternative to cover the interacting case and give a qualitative account of the behavior of the interacting non-commutative inflation. This prescription is essentially an application of an important Von Neumann theorem about the general representation of Von Neumann algebras and allows us to define a new Hamiltonian for non-commutative inflation that is indeed a time-translation generator of the very same algebra of generators that defines the dispersion relation Casimir. That is, allow us to deal only with $\pi_{\m{A}}$ representations. Besides that, it assures to recover the corresponding Poincaré representation at low energy and momentum. The new Hamiltonian (\ref{NewH}) is a function not only of the photon's dispersion relation but a function of the dispersion relation of all other particles of the theory.

To obtain the new Hamiltonian for non-commutative inflation, we need the direct integral decomposition of the representation (\ref{Fock}), that is a solved problem in the literature since the early days of QFT (for example, \cite{Shirokov}, \cite{ShirokovII}). We will address the problem of obtaining the new equations of state in a future paper. The main consequence of our results is that all the work previously done on non-commutative inflation must be redone, that is, finding a new equation of state, finding inflationary dispersion relations that assure successful inflation, etc. In particular, using the same class of dispersion relations of the original equations of the model will imply the limitation of the momentum of the theory, even the macroscopic one. We must, however, use the same rule to construct the Hamiltonian even if our theory does not have that limitation of momentum, otherwise we would fall on the very same problem discussed in this paper.

\begin{acknowledgements}
\C{U.D.M. thanks the Brazilian agency CNPq (142393/2006-1) and (246020/2012-1) for financial support. R.O. thanks FAPESP (06/56213-9) and the Brazilian agency CNPq (300414/82-0) for partial support.} 
\end{acknowledgements}


\end{document}